\documentclass[sigconf]{acmart}

\AtBeginDocument{%
  }

\copyrightyear{2025}
\acmYear{2025}
\setcopyright{acmlicensed}
\acmConference[WSDM '25]{Proceedings of the Eighteenth ACM International Conference on Web Search and Data Mining}{March 10--14, 2025}{Hannover, Germany}
\acmBooktitle{Proceedings of the Eighteenth ACM International Conference on Web Search and Data Mining (WSDM '25), March 10--14, 2025, Hannover, Germany}
\acmDOI{10.1145/3701551.3703521}
\acmISBN{979-8-4007-1329-3/25/03}

\usepackage{algorithmic}
\usepackage[ruled,linesnumbered]{algorithm2e}
\usepackage{siunitx}
\usepackage{enumitem}
\usepackage{multirow}
\usepackage{subcaption}

\begin{document}

\title[Tackling Temporality and Scalability with an Incremental Approach with Metropolis-Hastings Random Walks]{Optimizing Blockchain Analysis: Tackling Temporality and Scalability with an Incremental Approach with Metropolis-Hastings Random Walks}

\author{Junliang Luo}
\orcid{0009-0003-6921-6918}
\affiliation{%
  \institution{McGill University}
  \city{Montreal}
  \state{Quebec}
  \country{Canada}
}
\email{junliang.luo@mail.mcgill.ca}

\author{Xue Liu}
\orcid{0000-0001-5252-3442}
\affiliation{%
  \institution{McGill University}
  \city{Montreal}
  \state{Quebec}
  \country{Canada}
}
\email{xueliu@cs.mcgill.ca}

\renewcommand{\shortauthors}{Junliang Luo and Xue Liu}

\begin{abstract}
Blockchain technology, with implications in the financial domain, offers data in the form of large-scale transaction networks.
Analyzing transaction networks facilitates fraud detection, market analysis, and supports government regulation.
Despite many graph representation learning methods for transaction network analysis, we pinpoint two salient limitations that merit more investigation.
Existing methods predominantly focus on the snapshots of transaction networks, sidelining the evolving nature of blockchain transaction networks.
Existing methodologies may not sufficiently emphasize efficient, incremental learning capabilities, which are essential for addressing the scalability challenges in ever-expanding large-scale transaction networks.
To address these challenges, we employed an incremental approach for random walk-based node representation learning in transaction networks.
Further, we proposed a Metropolis-Hastings-based random walk mechanism for improved efficiency.
The empirical evaluation conducted on blockchain transaction datasets reveals comparable performance in node classification tasks while reducing computational overhead.
Potential applications include transaction network monitoring, the efficient classification of blockchain addresses for fraud detection or the identification of specialized address types within the network.
\end{abstract}




\maketitle

\begin{figure*}[htbp!]
\centering
\includegraphics[width=0.9\linewidth]{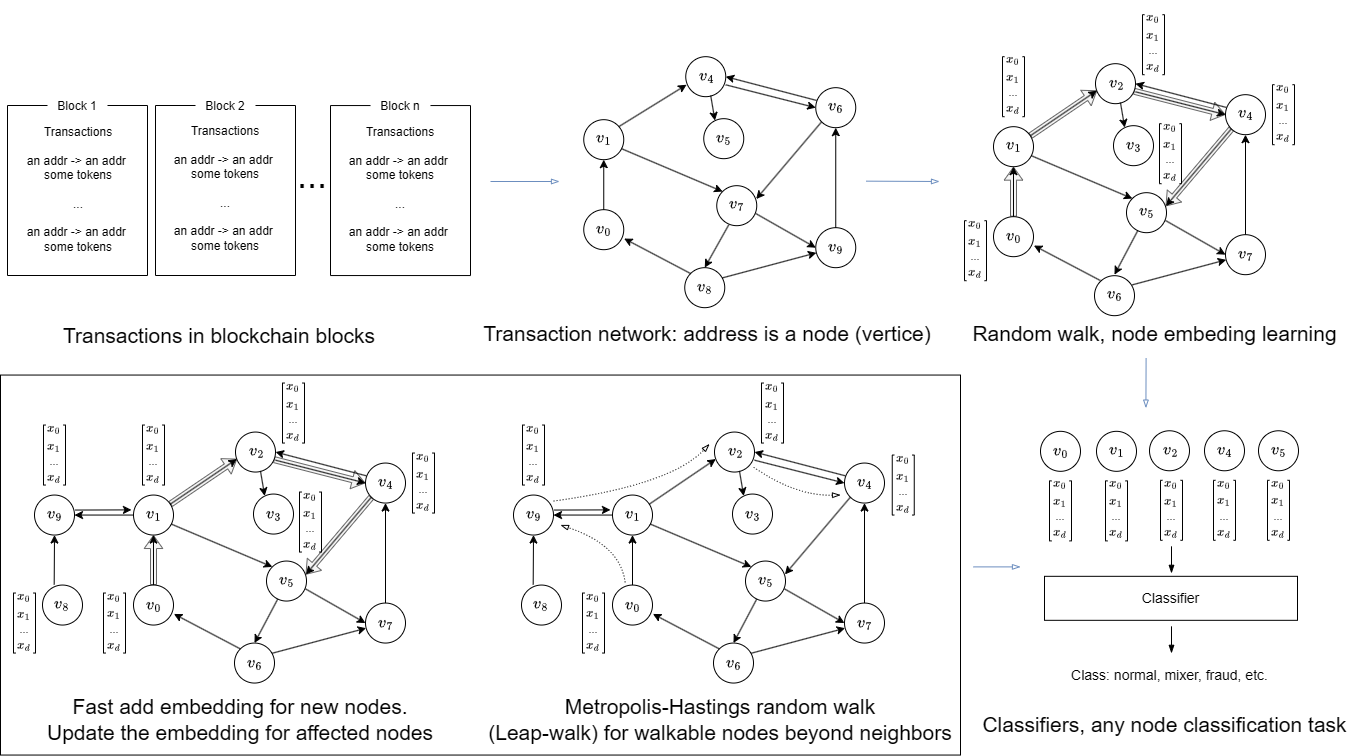}
\caption{Overview of the address (node) classification pipeline in a blockchain transaction network. Starting from raw transactions in blockchain blocks transformed into a transaction graph. Random walk method is performed to generate node embeddings. Our method (illustrated in the rectangular container) is towards addressing the challenges of  incremental learning and scalable solution of optimizing walk efficiency.}
\label{fig:transaction_random_walk}
\end{figure*}

\section{Introduction}
Blockchains use tamper-evident decentralized ledgers to store all transactions \cite{blockchain_overview}. 
This ledger is disseminated across multiple machines within the blockchain infrastructure, ensuring that the record is publicly accessible, transparent and resilient to tampering \cite{min2019blockchain}.
Transactions stored in a blockchain can be modeled as a transaction network (graph) representing the interaction (e.g. cryptocurrency transfers) among the network's participants.
These participants are essentially represented by addresses, modeled as graph nodes, with directed edges indicating the directional transfer of assets.
Academically, identifying distinct entities in blockchain transaction graphs is gaining attention as the potential to yield insights into transactional patterns \cite{khan2022graph}. 
Advancements in this research field carry promising implications for cybersecurity, legal compliance and governmental regulation \cite{sun2019regulating}.
This identification enables enhanced tracking and analysis of transactional behaviors, mitigating risks associated with fraudulent activities \cite{di2017blockchain}, and assisting the establishment of decentralized reputation mechanisms, reinforcing reliability in the decentralized financial systems \cite{harlev2018breaking}.
In addressing the challenge of entity identification in transaction graphs, approaches have been proposed based on node representation learning and machine learning classifiers, given their proven performance in networks such as citation, and social networks \cite{zhou2023cryptocurrency}.
The entities of nodes (addresses) in transaction graphs encompass those associated with illicit, fraudulent, and phishing activities \cite{alarab2020competence, kumar2020detecting, wu2020phishers, li2022ttagn}, mixers that obscure the origins of a cryptocurrency transaction for untraceability \cite{maksutov2019detection, wu2021detecting}, particular communities for market analysis \cite{ao2021temporal, chen2019market}, etc.
In terms of the methodologies, existing studies such as \cite{wu2020phishers, wu2021detecting, yuan2020detecting, patel2020graph, alarab2020competence, kudo2020gcnext}, emphasized predominantly static transaction graphs. These studies analyzed the full transaction graph over a specific time window.
Static graph methods neglect the temporal dynamics present in real-world blockchain transaction graphs as the graphs continually grow over time.
Research works including \cite{cai2021structural, pareja2020evolvegcn}, focused on dynamic graphs  that represent nodes and edges as temporal instances across various graph snapshots. 
The primary objective of those proposed methods is to detect node behaviours or forecast edge transactions leveraging these temporal features.
The studies were limited and seldom targeted the growing nature of blockchain transaction networks and scalability \cite{zhou2023cryptocurrency}.
Transaction networks from active blockchain projects consistently exhibit the emergence of new nodes and edges from ongoing transactions \cite{ron2013quantitative}.
This nature brings two primary issues for linking the previously proposed methods to the successful deployment on large-scale real-world transactions: the computational overhead of incremental learning and the magnitude of the expansive quantity.
The majority of recent methods proposed need to re-generate vector representations for the entire graph again upon any change \cite{zhou2023cryptocurrency}, and both static and dynamic approaches lack discussing reduced computation cost on large-scale graphs.
Towards addressing the challenges, a node representation learning algorithm that can update vector representations without full retraining on the entire graph again and reduces the computational overhead is needed.
Following the approaches of previous studies \cite{wu2020phishers, beres2021blockchain, lin2020t, 10174907, martin2022anomaly, wang2022heterogeneous}, we proposed a random-walk-based approach for its parallelism ability across nodes, offering improved efficiency potentials.
We first incorporated the strategy of Unbiased Update proposed by Sajjad et al. \cite{sajjad17icient} to adjust node representations without full graph retraining by trimming random walks at affected nodes and then continuing the walks, considering the graph's updated structure.
Then we proposed a Metropolis-Hastings random walks method that integrates leap-walk exploration, allowing walks to span beyond direct neighbors. 
This method makes the node transitions within walks reflect the desired importance of the nodes.
This mechanism resembles selectively sampling $h$-hop neighbors, guiding the leap-walks to obtain the embeddings with comparable classification performance with fewer walks, reducing the computational overhead.
We conducted experiments using four transaction networks including Bitcoin, Ethereum, and Ethereum Classic (ETC).
The experimental results show that the Unbiased Update altered the walks in a manner identical to regenerating the entire graph.
Furthermore, the proposed Metropolis-Hastings random walk methods achieved comparable F1 scores across the aforementioned datasets in a node classification task, using fewer walks than various random-walk-based comparison methods.
%
%
Lastly, we discussed the potential applications of the proposed methods, including transaction monitoring and smart contract mapping, and the limitation of the methods.

\section{Preliminary}
In blockchain transaction analytics with machine learning, graph representations offer a tool for understanding the intricate interactions within the network. 
This section describes the basics, covering the formulation of blockchain transaction graphs and random walk-based node representation learning. 

\subsection{Blockchain Transaction Networks}
A blockchain can be expressed as $\mathcal{B} = \{B_1, B_2, \dots, B_n\}$ ordered chronologically with the newest block appended at the end, where each block $B_i$ contains a set of transactions $T_i = \{t_1, t_2, \dots, t_m\}$.
Each transaction $t_k \in T_i$ has a sender $s_k$ and a receiver $r_k$.
From a segment $\mathcal{B}_{i }$ (where $1 \leq i \leq j \leq n$), we construct a graph $G = (V, E)$ from the union of its transactions.
$V$ is the set of participants, $V = \{s_k, r_k | t_k \in \bigcup_{l=i}^{j} T_l\}$, and $E \subseteq \bigcup_{k=i}^{j} T_k$ symbolizes the transaction interactions among them. 
$G$ encapsulates the transactions in the blocks $B_i$ to $B_j$ into a static graph.

\subsection{Node Representation Learning}
Given a graph G, characterized by a set of nodes $V = \{v_1, ..., v_n\}$ and a set of edges $E = \{e_1, ..., e_m\}$, where an edge $e_k$ between nodes $v_a$ and $v_b$ could carry a direction and a weight, notated as $e_k(v_a, v_b, w_k)$ representing the source node, target node, and the weight.
Node representation learning method $f$ projects nodes $V$ into a low-dimensional ($d$) latent space: $f: V \rightarrow \mathbb{R}^{d}, \text{ where } d \ll |V|.$ resulting in node embedding vectors 
$z_{i} = f(v_{i}) \in \mathbb{R}^{d}, \quad \forall v_{i} \in V$.
The geometric relationships between these vectors in the latent space echo the inherent relational present in the original graph \cite{hoff2002latent}.
From the produced vectors, features existing among the nodes can be captured and processed by statistical or machine learning models, enabling the modeling of graph-based data.

\subsubsection{Inner-product Methods}

\

\noindent
Node representation learning can be viewed from an encoder-decoder perspective, where the encoder projects nodes into a latent space to generate node embedding vectors, and the decoder utilizes the generated vectors to recreate or predict characteristics or relationships in the graph \cite{hamilton2020graph_encoder_decoder}.
The encoder-decoder framework fosters the feasibility of optimizing embeddings, facilitating a more refined correlation between the distances in the embedding space and the relative positioning of nodes within the original graph topology through learning.
Inner-product methods, built upon the Laplacian eigenmaps technique \cite{belkin2001laplacian}, employ an inner-product-based decoder defined as $\text{dec}(z_u, z_v) =  z_u^\top z_v$. 
This decoder design implies that the proximity in the embedded space, as measured by the inner product, reflects the similarity in the original graph.
The loss to be minimized can be defined as, for example, in \cite{hamilton2020graph_inner_product}: $L = \sum_{(u,v)\in D} \left||\text{dec}(z_u, z_v) - S[u, v]\right||^{2}_{2}$, where $S[u, v]$ is a node-level similarity function to be defined.

\subsubsection{Random Walks}

\

\noindent
$S[u, v]$, the node-node similarity can be based on the stochastic
measures of neighbourhood overlap.
Node embeddings can be optimized to encourage similarity between pairs of nodes if they exhibit a propensity to co-occur on short random walks over a graph.
The nodes sharing similar random walk patterns are likely to share similar structural roles within the graph, thus should be mapped close to each other in the embedding space.
Consider a random walk $W = \{w_1, w_2, ..., w_L\}$ of length $L$. For each node $w_i$ in the walk, a context window of size $p$ is defined to extract node pairs that are within the window. 
The node pairs are then used to learn the node representations in a model such as skip-gram \cite{mikolov2013efficient}.
Examples of methods utilizing random walks for node representation learning include DeepWalk \cite{perozzi2014deepwalk}: uniform random walks for community detection, node2vec \cite{grover2016node2vec}: biased walks for flexibility in exploring local and global structures, and LINE \cite{tang2015line}: preserves first- and second-order proximities.

\subsubsection{Skip-Gram Model}

\

\noindent
The Skip-Gram model enables the actual learning of the node embeddings by leveraging the co-occurrence patterns of nodes in short random walks.
The encoder maps each node to its corresponding embedding vector and the decoder predicts the context nodes that tend to co-occur in the random walks given the embedding vector of an input target node, as proposed in \cite{mikolov2013efficient}.
Let $N(u)$ denote the set of context nodes associated with a node $u$ in the random walks. 
The encoder function is $f(u) = z_u$, where $z_u$ is the embedding vector.
The similarity between the target node \(u\) and a context node \(v\) can be measured by the probability $p_\mathcal{G}(v|u)$: the likelihood of node $v$ being visited from node $u$ during a random walk.
The conditional probability $p_\mathcal{G}(v|u)$ is often modeled using a softmax function, which takes the inner product of the embeddings of $u$ and $v$, and normalizes it over all possible nodes $V$ in the graph.
\begin{equation}
    \text{Dec}(z_u, z_v) = .
    \frac{{e^{(z_u^\top z_v)}}}{{\sum_{k=1}^{|V|} e^{(z_u^\top z_k)}}}
    \simeq p_\mathcal{G}(v|u)
\end{equation}
The loss function is formulated based on the negative log-likelihood  defined as $L = \sum_{u,v \in T}-\log\left(\text{Dec}(z_u, z_v)\right)$, where $T$ is the training set of random walks.
Minimizing the loss is to learn embedding vectors that maximize the likelihood of observing the actual context nodes given the target nodes in the random walks.

\begin{table*}
    \centering
    \caption{Comparison of MAE of Empirical and Theoretical Transition Probabilities.}
    \label{tab:mae_}
    \resizebox{0.94\textwidth}{!}{%
     \setlength{\tabcolsep}{10pt} 
    \begin{tabular}{l *{10}{S[table-format=1.4]}}
        \toprule
        Method & \multicolumn{10}{c}{Percentage of Total Transactions (Initial + 5\% Increments)} \\
        \cmidrule(lr){2-11}
        & {50+5} & {55+5} & {60+5} & {65+5} & {70+5} & {75+5} & {80+5} & {85+5} & {90+5} & {95+5} \\
        \midrule
        From Scratch & 0.0541 & 0.0550 & 0.0559 & 0.0557 & 0.0552 & 0.0555 & 0.0551 & 0.0550 & 0.0546 & 0.0548 \\
        
        Unbiased Update & 0.0545 & 0.0546 & 0.0559 & 0.0559 & 0.0552 & 0.0555 & 0.0553 & 0.0550 & 0.0548 & 0.0549 \\
        
        Naive & 0.0659 & 0.0611 & 0.0622 & 0.0610 & 0.0604 & 0.0606 & 0.0596 & 0.0580 & 0.0580 & 0.0571 \\
        
        \midrule
        \emph{Diff. (Scratch vs Unbiased)} & \num{0.0004} & \num{0.0004} & \num{0.0000} & \num{0.0002} & \num{0.0000} & \num{0.0000} & \num{0.0002} & \num{0.0000} & \num{0.0002} & \num{0.0001} \\
        
        \emph{Diff. (Scratch vs Naive)} & \num{0.0118} & \num{0.0061} & \num{0.0063} & \num{0.0053} & \num{0.0052} & \num{0.0051} & \num{0.0045} & \num{0.0030} & \num{0.0034} & \num{0.0023} \\
        \bottomrule
    \end{tabular}
    }
\end{table*}

\section{Incremental learning}
It is imperative to extend the existing static random walk-based node representation methods proposed for blockchain transaction graphs (e.g. a substantial number covered by the review \cite{zhou2023cryptocurrency}) to be with efficient incremental learning capabilities. 
A general strategy that is compatible with various existing random walk-based methods, is worth discussing.
Such a strategy should be able to efficiently update the vertex representations after changes in the graph without needing to re-train a model from scratch.
Further, what demands emphasis is that the strategy must not introduce biases in the updated random walks.

\subsection{Random Walk Unbiased Update}

Drawing inspiration from and adapting the strategy Unbiased Update proposed by Sajjad et al. \cite{sajjad17icient}, 
the strategy includes generating new walks for newly added nodes, identifying previously existing nodes impacted by the changes, and refining the random walks associated with these nodes.
Specifically, for any walk that encounters an affected node in the current graph, the algorithm trims the walk back to the first occurrence of the affected vertex and then continues the walk, ensuring it considers the updated graph, therefore enables the incremental learning without introducing bias.

\begin{algorithm}
\caption{Unbiased Update \cite{sajjad17icient} for Transaction Graph}
\label{algo:unbiased_blockchain}
\begin{algorithmic}[1]
\REQUIRE $G_t(V_t, E_t)$, $G_{t+1}(V_{t+1}, E_{t+1})$, $num\_walks$, $walk\_length$
\ENSURE Updated walks $W_{t+1}$
\STATE $\Delta V \gets \{ v \mid \exists (v, u) \in E_{t+1} \ \text{and} \ (v, u) \notin E_t \}$
\STATE $V_{n} \gets V_{t+1} \setminus V_t$
\STATE $V_a \gets V_t \cap \Delta V$
\STATE $W_{a} \gets \{ w \mid w \in W_t, \exists v \in V_a: v \in w \}$
\FOR{each $w \in W_{a}$}
    \STATE  $w^* \gets $ trim $w$ at first $v \in V_a$, yielding $w^*$
    \STATE $w^+ \gets $ resample $w^*$ starting from last node in $w^*$ in $G_{t+1}$
    \STATE $W_{re} \gets $ append $w^+$
\ENDFOR
\STATE $W_{n} \gets$ generate new walks for each $v \in V_{n}$ in $G_{t+1}$
\STATE $W_{t+1} \gets (W_t \setminus W_a) \cup W_{re} \cup W_{n}$
\STATE Use $W_{t+1}$ to retrain the model
\end{algorithmic}
\end{algorithm}

The Unbiased Update for transaction graphs is presented in Algorithm. \ref{algo:unbiased_blockchain}. 
$\Delta V$ captures the nodes with changed connectivity between sequential graph versions. 
$V_{n}$ and $V_a$ represent the new and affected nodes, respectively. 
Affected walks $W_a$ include vertices in $V_a$. These walks are trimmed and resampled, integrating changes from $G_{t+1}$ to produce the updated walk set $W_{t+1}$ used for model retraining.
Compared to the original Unbiased Update algorithm from \cite{sajjad17icient}, we make two changes for blockchain transaction data:
One, affected nodes stem solely from new transactions, as blockchain’s immutability precludes edge removals. 
Two, the algorithm integrates new transactions ($W_{n}$) into the walk set to capture evolving transactions network relevance.
Algorithm. \ref{algo:unbiased_blockchain} updates the model by identifying the changes in the graph between time instances $t$ and $t+1$. 
Identifying the affected nodes and edges has a time complexity of $O(|E_{t+1}|)$. The trimming of affected walks has a time complexity of $O(|W_a| \times \textit{walk\_length})$ and resampling the walks necessitates $O(|W_a| \times \textit{walk\_length})$. 
Even when summed, these complexities are considerably less than the $O(|V_{t+1}| \times \textit{num\_walks} \times \textit{walk\_length})$ time complexity of the retrain the whole new transaction graph from scratch. 
Particularly in real-world scenarios where $|V_a| + |V_n| \ll |V_{t+1}|$, thus the method provides significant computational savings compared to retraining the whole graph from scratch.
The strategy preserves historical context (for the unaffected nodes, the associated walks keep unchanged), and integrate also the learning of walks of new nodes and update the walks of affected nodes unbiasedly. 
%
%
Additionally, the strategy is compatible with random walk-based graph methods, to have them be with complexity scaling linearly by affected vertices and edges, enabling the use on large-scale blockchain graphs.

\subsection{Evaluation of Unbiased Update}
The evaluation of Algorithm \ref{algo:unbiased_blockchain} is the same as suggested in \cite{sajjad17icient} using a comparative 
evaluation metric derived from random walks. 
Walks are generated on a new graph from scratch, as opposed to updating walks trained on a previous state graph to reflect changes in the new state graph at each timestamp.
The evaluation metric employed is the MAE $\delta$ between empirical 
transition probabilities and the theoretical transition probabilities.
$\delta = \frac{1}{N} \sum_{N}^{} |\hat{P}(v_i, v_j) - P(v_i, v_j)|$
, where $\hat{P}(v_i, v_j)$ and $P(v_i, v_j)$ denote the empirical and theoretical transition probabilities from $v_i$ to $v_j$, respectively, where N is represents the total number of transitions (pairs of neighbour vertices).
Empirical transition probability refers to the transition probability observed in the generated random walks.
Theoretical transition probability is the underlying transition probability obtained from the whole graph, calculated as the inverse of the node's degree.
The dataset employed in our study is the Ethereum transaction dataset collected in \cite{wu2020phishers}, which encompasses \textit{2.9} million addresses and over 13 million transactions.
The dataset was partitioned to simulate the temporal evolution of the transaction graph.
We initially sorted all the transactions by date ascending, and divided the dataset into two halves and for the second half subsequently make segments increasing by 5\% of the total transactions.
We then construct a series of graphs from the segments reflecting time-ordered transactions. 
Each successive graph encapsulates all prior data.
In the experiments, the From Scratch method trains on each graph individually, while the Unbiased Update approach incrementally trains on the previous graph and quickly retrains on each new graph with 5\% added transactions using the algorithm.
A Naive update for comparison is upon encountering new nodes in an evolving graph, only generate walks for new added nodes while retaining the walks from the previous graph unchanged.
The results presented in Table \ref{tab:mae_} show that the Unbiased Update generates walks much closer to being statistically indistinguishable from those of the From Scratch method, compared to the Naive approach.

\section{Metropolis-Hastings Random Walk}
The evolving nature of blockchain transaction networks poses challenges in handling new data and efficiently processing their growing scale.
To effectively apply machine learning models for detecting various entities within the growing scale of blockchain transactions, it is essential to improve computational efficiency, enhance representation learning effectiveness, and conduct thorough examination of transaction networks \cite{taherdoost2023blockchain}.
Given that some nodes of certain characteristics in the blockchain transaction graph can be considered as less relevant when detecting certain identity or community,  sampling strategies can be applied to reduce the computation cost.
Sampling strategies on the transaction graph tackle putting computational resources on the more informative portions of the graph, reducing the cost of graph learning and potentially also improving the quality of the learned representations \cite{zeng2019graphsaint}.
We proposed a random walk-based method that resembles sampling strategies based on the Metropolis-Hastings (MH) algorithm \cite{metropolis1953equation, chib1995understanding}.
The approach can be visualized as a leap-walk, where certain nodes within $h$-hop neighbours are more likely to be sampled than others.
The Metropolis-Hastings algorithm is from the family of Markov chain Monte Carlo (MCMC) methods \cite{brooks1998markov}. 
Let $\mathcal{X}$ denote the state space of a Markov chain with transition matrix $\mathbf{P}$. 
MCMC methods generate a sequence of random samples, $X_1, X_2, \ldots, X_T$ from a desired probability distribution $p^*$ over $\mathcal{X}$. 
This is achieved by constructing a Markov chain with stationary distribution $p$, ensuring that, as $T \to \infty$, the generated sequence converges to $p$.
A crucial condition for the convergence of the Markov chain to the stationary (target) distribution $p$ is the detailed balance condition:
\begin{equation}
p(x) \mathbf{P}(x, x') = p(x') \mathbf{P}(x', x) \quad \forall x, x' \in \mathcal{X}
\label{detailed_balance_condition}
\end{equation}
In the Metropolis-Hastings method, given a current state $x \in \mathcal{X} $, a candidate next state $x'$ is proposed from a known, often symmetric, proposal distribution $q(x' | x)$. 
\begin{equation}
R(x, x') = \frac{p(x') q(x | x')}{p(x) q(x' | x)}
\label{MH_Equation}
\end{equation}
In Equation. \ref{detailed_balance_condition}, $P$ denotes the inherent transition probabilities of the Markov chain, reflecting the systematic behavior of state transitions. Complementing this, $q$ acts as a proposal mechanism, suggesting potential transitions between states based on specific criteria.
The ratio of target densities from state $x$ to $x'$ is denoted by $R(x, x')$:
Thus, the acceptance probability is at most one expressed as:
\begin{equation}
\alpha(x, x') = \min\left(1, R(x, x')\right)
\end{equation}
If the proposal is accepted with probability $\alpha$, $x'$ becomes the next state in the chain; otherwise, the chain remains in state $x$.
For blockchain transaction graphs, the Metropolis-Hastings algorithm is able to guides the sampling process through the transnational relations. 
In the blockchain transaction graph context, $p$ represents the target distribution that encapsulates the transactional importance or relevance of  addresses. 
The proposal distribution $q$ guides the traverse in the transaction graph.
By proposing new states based on $q$, the algorithm promotes thorough exploration of the network’s expanse.
The algorithm balances its focus between regions of significant transactional activity and less-explored areas of the blockchain transaction graph.
This is achieved through choosing the proper distribution $p$ and $q$.
Table. \ref{tab:functions} presents the potential functions for the target distribution $p$ and the proposal distribution $q$ for blockchain transaction graphs.

\begin{table}[t!]
    \centering
    \caption{ $p$ and $q$ functions for blockchain transaction data.}
    \label{tab:functions}
    \resizebox{0.466\textwidth}{!}{%
    \begin{tabular}{ll}
        \hline
        \textbf{Function} & \textbf{Description} \\
        \hline
        $p_{V_{in}}(u)$ & Total incoming transaction value of node $u$ \\
        $p_{V_{out}}(u)$ & Total outgoing transaction value of node $u$ \\
        $p_{F}(u)$ & Frequency of transactions of node $u$ \\
        $p_{D_{in}}(u)$ & Degree of incoming transactions for node $u$ \\
        $p_{D_{out}}(u)$ & Degree of outgoing transactions for node $u$ \\
        $q_{S}(u, v)$ & Reciprocal of the shortest distance (hop) from $u$ to $v$ \\
        $q_{E}(u, v)$ & Exponential decay based on the distance from $u$ to $v$: \\ & $e^{-\lambda \times \text{dist}(u, v)}$, \\
        \hline
    \end{tabular}}
\end{table}

%
%
%
The target (stationary) distribution $p$ encapsulates specific characteristics of the nodes, seen as the importance of addresses based on their transactional activity. 
The proposal distribution $q$ recommends the transitions to nearby addresses, primarily influenced by their distance from the current address.
The progression of the walk from one node to another mirrors the transition between states in the Metropolis-Hastings algorithm: the transition probability between node $u$ and $v$ in the random walks will be: $\alpha(u, v) = \min\left(1, R(u, v)\right)$.
The transition probability  from an address node $u$ to $v$ is a ratio: $ p(v)q(u|v) / p(u)q(v|u) $ same as in Equation. \ref{MH_Equation}.
%

%
%
%

\begin{algorithm}[t!]
\caption{Metropolis-Hastings Random Walks Sampling}
\label{algo:MH_sampling}

\KwData{Graph $G(V, E)$, number of walks $n$, walk length $l$, func $p()$, $q()$, hops $h$, min acceptance threshold $\alpha_{\text{m}}$}

\KwResult{Walks $W$}

$W \leftarrow \emptyset$\;
\ForEach{$u \in V$}{
    \For{$i \leftarrow 1$ \KwTo $n$}{
        $w \leftarrow [u]$\;
        $curr \leftarrow u$\;
        \For{$j \leftarrow 1$ \KwTo $l - 1$}{
            $N_h(curr) \leftarrow \{v \mid dist(curr, v) = h \}$\; 
            $v \sim \text{Uniform}(v|N_h(curr))$;  \enspace   
            $\alpha \leftarrow \min\left(1, \frac{p(v)q(curr|v)}{p(curr)q(v|curr)}\right)$\;
            Draw $r \sim \text{Uniform}(0, 1)$\;
            \If{$r < \alpha + \alpha_{m}$}{
                Append $v$ to walk $w$ and set $curr \leftarrow v$\;
            }
        }
        Append walk $w$ to list $W$\;
    }
}
\Return{$W$}\;
\end{algorithm}

\begin{figure}[htbp!]
    \centering
    \includegraphics[width=0.36\textwidth, scale=1]{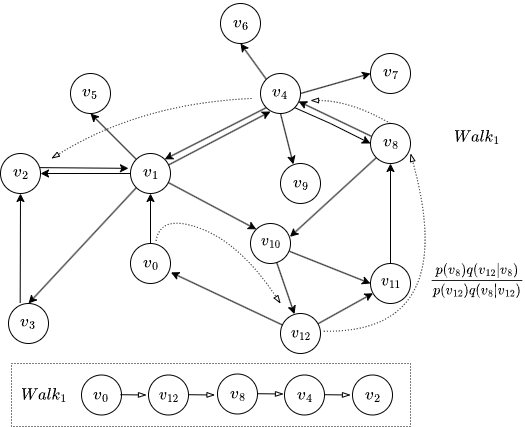} 
    \caption{Illustration of the Metropolis-Hastings Random Walk (Leap-walk) on a transaction graph, where nodes represent addresses and edges represent token transfers. The highlighted dotted curved pathway exemplifies that leap-walk exploration enables sampling beyond immediate neighbors.}
    \label{fig:leak_random_walk}
\end{figure}

Algorithm \ref{algo:MH_sampling} describes the details of the proposed random walk-based algorithm. 
In each step, instead of moving to immediate neighbors, a node is sampled uniformly from nodes that are exactly $h$-hops away. 
An acceptance probability $\alpha$ is then computed, which will be summed with a predefined minimum threshold $\alpha_{\text{m}}$, dictates the likelihood of the walk transitioning to the selected node. 
As the transaction graph is directed, $N_h(curr)$ guarantees that $v$ reachable for denominator in MH has a value for $q(v|curr)$. 
The numerator $q(curr|v)$ depends on the node $curr$'s direct reachability from $v$. If $v$ is not returnable, $q(curr|v)$ will be set a nominal probability of .1.
After generating the Metropolis-Hastings random walks, we then train a model to learn an embedding vector for each node. 
Similar to DeepWalk \cite{perozzi2014deepwalk}, we utilize the skip-gram model to maximize the probability of co-occurrence of nodes within a context window. 
Let $W$ represent the set of all random walks generated by Algorithm \ref{algo:MH_sampling}. For each walk $w \in W$, let $v_i$ be a node in $w$, and consider a context window of size $c$ around $v_i$. 
The objective is to maximize the probability of observing the context nodes $v_{i-j}$ and $v_{i+j}$, where $j \in \{1, \dots, c\}$, given the current node $v_i$. The training objective is defined as follows: \[ \max \sum_{w \in W} \sum_{v_i \in w} \sum_{-c \leq j \leq c, j \neq 0} \log \Pr(v_{i+j} | v_i) \]
The probability $\Pr(v_{i+j} | v_i)$ is computed using the softmax function: \[ \Pr(v_{i+j} | v_i) = \frac{\exp(\mathbf{z}_{v_{i+j}} \cdot \mathbf{z}_{v_i})}{\sum_{v \in V} \exp(\mathbf{z}_v \cdot \mathbf{z}_{v_i})} \] where $\mathbf{z}_v$ is the embedding vector for node $v$ in the graph. 
By optimizing this objective function, we learn embedding vectors $\mathbf{z}_v$ for each node $v \in V$.
We propose the adoption of the Metropolis-Hastings based sampling strategy for random walks on blockchain transaction graphs as it introduces two advantages:
\begin{enumerate}
    \item \textit{Balancing Mechanism}: The transition probability from $u$ to $v$ is decided by the ratio considering both $q(u|v)$ and $q(v|u)$ to acts as a balancing mechanism.
    This mechanism corrects the biased traversal across nodes, ensuring the sampling process isn't one-sided or biased towards a particular traversal direction, i.e., both outgoing and incoming transactional relationships of a node are taken into account.
    
    \item \textit{Leap-walk Exploration}: This approach facilitates the exploration that extends beyond the immediate neighbors, enabling walks to transition to nodes situated at a distance of more than 1-hop. As Figure. \ref{fig:leak_random_walk} demonstrated, the mechanism we term as leap-walk.
    The node representations are anticipated to be obtained with a reduced number of walks, as the method prioritizes producing embedding vector for nodes from fewer, high-utility walks, achieving competitive performance for downstream tasks.
\end{enumerate}

\subsection{Evaluation of Metropolis-Hastings Random Walk}
To evaluate the proposed Metropolis-Hastings random walks,  we discuss two aspects: the performance of the produced node embedding on downstream node classification tasks, and the cost mainly measured by the quantity of walks required and the resulting performance.
We benchmark the performance by focusing on the discriminative ability of using the derived embeddings in distinguishing various node categories with common machine learning classifiers: logistic regression (LR) and support vector machine (SVM). 
Additionally, since MH and the subsequent baseline and comparison algorithms are all based on the skip-gram approach, we consider the quantity of random walks a crucial metric for evaluating computational cost.

\subsubsection{Datasets}
\begin{description}[leftmargin=0pt, itemsep=0.5em, font=\normalfont\itshape]

\item[Ethereum Transaction \cite{xblockEthereum}]
This dataset presents a transaction graph that encompasses 2,973,489 nodes and 13,551,303 edges. Out of these nodes, 1,165 are labeled as illicit while the remaining nodes are unlabeled. The graph is of an average degree of 4.5574.

\item[Bitcoin Elliptic Transaction \cite{weber2019anti}]
This dataset offers a transaction graph where nodes depict transactions and the edges represent the flow of Bitcoins between these transactions. The graph comprises 203,769 nodes and 234,355 edges with 4,545 nodes labeled illicit, 42,019 are labeled licit, and the remainder are unlabeled.

\item[Phishing Node Dataset \cite{lin2020t}]
The dataset comprises Ethereum transactions sourced from 445 identified phishing addresses, paired with an equal number of random unlabeled addresses. Through K-order sampling (with \( K_{\text{in}} = 1 \) and \( K_{\text{out}} = 3 \)) for each of the 890 objective nodes, subgraphs were collected and then merged into an expansive network that includes 86,623 nodes.

\item[ETC Dataset \cite{al2020labeled}]
This transaction network dataset was crafted from Ethereum Classic (ETC) network. Initial data was sourced from Etherscamdb\footnote{https://github.com/MrLuit/EtherScamDB}. Anomalies were identified based on descriptions of scams, with redundant addresses filtered out. The datasets contains 73034 nodes with 2357 labeled as fake (scam), and 71,250 edges. 

\end{description}

\begin{table*}[!htbp]
\resizebox{0.999\textwidth}{!}{%
\begin{tabular}{llllllllllllllllllllllll}
\hline
 &  & \multicolumn{4}{l}{Ethereum} &  &  & \multicolumn{4}{l}{Bitcoin Elliptic} &  &  & \multicolumn{4}{l}{Phishing Node} &  &  & \multicolumn{4}{l}{ETC} \\
 &  & \multicolumn{2}{l}{LR} & \multicolumn{2}{l}{SVM} &  &  & \multicolumn{2}{l}{LR} & \multicolumn{2}{l}{SVM} &  &  & \multicolumn{2}{l}{LR} & \multicolumn{2}{l}{SVM} &  &  & \multicolumn{2}{l}{LR} & \multicolumn{2}{l}{SVM} \\ \cline{2-6} \cline{8-12} \cline{14-18} \cline{20-24} 
 & wn & Acc & F1 & Acc & F1 &  & wn & Acc & F1 & Acc & F1 &  & wn & Acc & F1 & Acc & F1 &  & wn & Acc & F1 & Acc & F1 \\
Node2vec & 1 & 0.637 & 0.617 & 0.678 & 0.559 &  & 5 & 0.616 & 0.662 & 0.630 & 0.691 &  & 5 & 0.702 & 0.683 & 0.753 & 0.732 &  & 5 & 0.515 & 0.582 & 0.529 & 0.501 \\
Walklets &  & 0.706 & 0.678 & 0.706 & 0.679 &  &  & 0.619 & 0.616 & 0.660 & 0.664 &  &  & 0.736 & \textbf{0.734} & 0.770 & 0.763 &  &  & 0.721 & 0.698 & 0.750 & 0.767 \\
Role2Vec &  & 0.818 & 0.805 & 0.833 & 0.819 &  &  & 0.652 & 0.663 & 0.726 & \textbf{0.714} &  &  & 0.770 & 0.745 & 0.719 & 0.742 &  &  & 0.721 & \textbf{0.740} & 0.751 & 0.773 \\
MH  $p_{V_{in}}(u)$, $q_{S}(u, v)$ &  & 0.835 & \textbf{0.813} & 0.841 & \textbf{0.823} &  &  &  &  &  &  &  &  & 0.685 & 0.705 & 0.775 & 0.780 &  &  & 0.779 & 0.737 & 0.809 & 0.764 \\
MH   $p_{D_{in}}(u)$, $q_{S}(u, v)$ &  & 0.826 & 0.801 & 0.837 & 0.814 &  &  & 0.646 & \textbf{0.676} & 0.677 & 0.703 &  &  & 0.719 & 0.709 & 0.770 & \textbf{0.781} &  &  & 0.765 & 0.714 & 0.794 & \textbf{0.791} \\
MH  $p_{V/D_{in}}(u)$, $q_{E}(u, v)$ &  & 0.82 & 0.790 & 0.841 & 0.822 &  &  & 0.644 & 0.672 & 0.675 & 0.701 &  &  & 0.713 & 0.705 & 0.753 & 0.75 &  &  & 0.751 & 0.712 & 0.824 & 0.786 \\
 &  &  &  &  &  &  &  &  &  &  &  &  &  &  &  &  &  &  &  &  &  &  &  \\
Node2vec & 2 & 0.841 & 0.824 & 0.861 & 0.850 &  & 10 & 0.641 & 0.675 & 0.713 & 0.763 &  & 8 & 0.770 & 0.776 & 0.775 & 0.780 &  & 10 & 0.662 & 0.676 & 0.721 & 0.678 \\
Walklets &  & 0.697 & 0.687 & 0.749 & 0.763 &  &  & 0.644 & 0.643 & 0.686 & 0.687 &  &  & 0.764 & 0.769 & 0.753 & 0.782 &  &  & 0.749 & 0.754 & 0.794 & 0.811 \\
Role2Vec &  & 0.863 & 0.858 & 0.886 & 0.883 &  &  & 0.661 & 0.674 & 0.771 & 0.780 &  &  & 0.781 & 0.782 & 0.725 & 0.759 &  &  & 0.794 & 0.829 & 0.765 & 0.795 \\
MH $p_{V_{in}}(u)$, $q_{S}(u, v)$ &  & 0.869 & \textbf{0.864} & 0.886 & \textbf{0.886} &  &  &  &  &  &  &  &  & 0.775 & \textbf{0.789} & 0.798 & \textbf{0.802} &  &  & 0.868 & \textbf{0.852} & 0.838 & 0.807 \\
MH $p_{D_{in}}(u)$, $q_{S}(u, v)$ &  & 0.856 & 0.844 & 0.871 & 0.865 &  &  & 0.69 & 0.706 & 0.75 & 0.774 &  &  & 0.77 & 0.776 & 0.792 & 0.801 &  &  & 0.853 & 0.839 & 0.868 & 0.847 \\
MH $p_{V/D_{in}}(u)$, $q_{E}(u, v)$ &  & 0.867 & 0.856 & 0.873 & 0.867 &  &  & 0.693 & \textbf{0.717} & 0.755 & \textbf{0.780} &  &  & 0.775 & 0.789 & 0.787 & 0.793 &  &  & 0.838 & 0.82 & 0.868 & \textbf{0.852} \\
 &  &  &  &  &  &  &  &  &  &  &  &  &  &  &  &  &  &  &  &  &  &  &  \\
Node2vec & 3 & 0.878 & 0.871 & 0.893 & 0.892 &  & 15 & 0.641 & 0.689 & 0.703 & 0.766 &  & 10 & 0.775 & 0.770 & 0.803 & 0.798 &  & 15 & 0.868 & 0.857 & 0.853 & 0.828 \\
Walklets &  & 0.672 & 0.662 & 0.766 & 0.765 &  &  & 0.653 & 0.650 & 0.716 & 0.706 &  &  & 0.787 & 0.784 & 0.787 & 0.789 &  &  & 0.824 & 0.829 & 0.838 & 0.845 \\
Role2Vec &  & 0.871 & 0.866 & 0.882 & 0.878 &  &  & 0.701 & 0.706 & 0.789 & 0.798 &  &  & 0.778 & 0.792 & 0.742 & 0.753 &  &  & 0.853 & 0.861 & 0.794 & 0.829 \\
MH $p_{V_{in}}(u)$, $q_{S}(u, v)$  &  & 0.893 & \textbf{0.886} & 0.895 & \textbf{0.893} &  &  &  &  &  &  &  &  & 0.787 & \textbf{0.796} & 0.792 & \textbf{0.806} &  &  & 0.914 & 0.912 & 0.897 & 0.889 \\
MH $p_{D_{in}}(u)$, $q_{S}(u, v)$ &  & 0.873 & 0.867 & 0.888 & 0.888 &  &  & 0.719 & 0.741 & 0.794 & \textbf{0.822} &  &  & 0.770 & 0.768 & 0.798 & 0.793 &  &  & 0.941 & \textbf{0.938} & 0.882 & 0.867 \\
MH $p_{V/D_{in}}(u)$, $q_{E}(u, v)$ &  & 0.876 & 0.868 & 0.893 & 0.891 &  &  & 0.722 & \textbf{0.744} & 0.79 & 0.819 &  &  & 0.775 & 0.780 & 0.803 & 0.804 &  &  & 0.897 & 0.892 & 0.912 & \textbf{0.909} \\
 &  &  &  &  &  &  &  &  &  &  &  &  &  &  &  &  &  &  &  &  &  &  &  \\ \hline
\end{tabular}
}
\caption{Performance evaluation of Metropolis-Hastings random walks compared with three algorithms: Node2vec, Role2vec, and Walklets. Three variants of MH: $p_{V_{in}}(u)$, $q_{S}(u, v)$; $p_{D_{in}}(u)$, $q_{S}(u, v)$; and $p_{V_{in}}(u)$ (or $p_{D_{in}}(u)$), $q_{E}(u, v)$, are presented due to better performance over other variants. The walk numbers (wn) represent performance saturation points, followed by deliberate reductions to evalute performance under constrained walk lengths, pertinent when walk length is limited for scalability. Accuracy, F1 score are as metrics and LR and SVM are as classifiers for node classification.}
\label{tab:mh_performance_comparison}
\end{table*}

\subsubsection{Evaluation Setup} 

\

\noindent In the experiment, we utilized the entire blockchain transaction graph as input to the Metropolis-Hastings method as well as the other comparison random-walk-based algorithms to generate the walks.
Subsequently, leveraging the skip-gram loss, these walks were utilized to unsupervisedly learn the embedding vector for each node.
Once the embedding vectors for each node were obtained, we trained LR and SVM classifiers. 
These classifiers used the node embeddings as the input, with ground truth labels from the datasets (e.g., illicit vs. licit, scam vs. normal) to perform supervise learning.
Given the limited number of positively labeled addresses, we sample multiple sets of negative address embedding vectors, each matching the number of positive address vectors. Each combined set, containing an equal number of positive and negative address vectors, serves as input for distinct datasets.
The classifier LR are configured with $saga$ solver, and the SVM with $rbf$ kernel, regularization $C$ of 10, and kernel coefficient $\gamma$ of .4, parameters chosen due to performance in preliminary tests.
For evaluation, we randomly split each dataset into training 80\% and testing 20\% subsets. The performance metrics employed include the mean accuracy and the F1 score.
For all the methods evalauted, we use a fixed walk length of 5, and same window size $c$ of 5, embedding dimension of 64, same as some early works on the same datasets \cite{wu2020phishers, 10174907}, a minimal feature occurrence count of 1, and a learning rate of .05.
For Metropolis-Hastings methods, we use a $h$ of 2-hop for the proposal distribution $q$ and $a_m=.5$.

\subsubsection{Benchmark Methods for Comparison}

\

\noindent
In the evaluation, we compared the MH methods with three other random walk-based algorithms:
Node2Vec \cite{grover2016node2vec}, a baseline algorithm for node embedding, leveraging random walks here we set $p=1$, $q=1$ to test on uniform transition probabilities across neighbors.
Walklets \cite{perozzi2017don}, an algorithm employs random walks to approximate the pointwise mutual information matrix derived from individual normalized adjacency matrix powers.
Role2Vec \cite{ahmed2018learning}, an algorithm employs random walks to estimate the pointwise mutual information matrix, which is derived by combining the pooled adjacency power matrix with a structural feature matrix (Weisfeiler-Lehman feature matrix).
The baseline algorithms are chosen since they are all random walk strategy and skip-gram model-based for learning node embeddings. 
The difference is that Walklets and Role2Vec introduce distinct matrix features, such as structural characteristics, enriching the features for training the Skip-gram model.

\begin{figure}[!htbp]
    \centering
    \includegraphics[width=0.46\textwidth, scale=1]{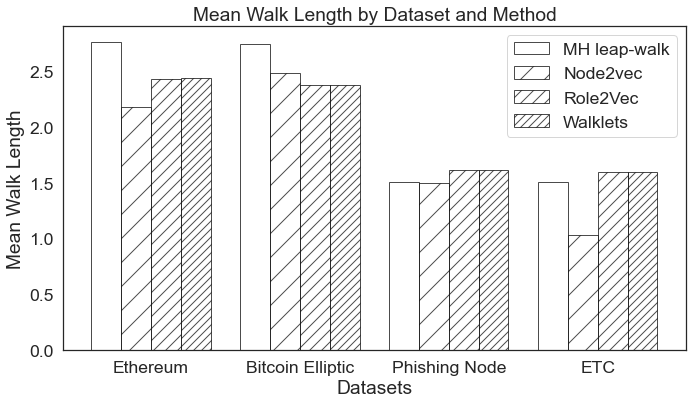} 
    \caption{Comparison of mean defacto walk lengths across methods for a fixed walk length of 5. MH methods consistently achieved a higher mean walk length compared to other methods, emphasizing its ability to traverse more unique nodes within the same walk length constraints.}
    \label{fig:mean_walk_length}
\end{figure}

\subsubsection{Experimental Results}

\

\noindent
As Table \ref{tab:mh_performance_comparison} presents, at walk number 1 on the Ethereum dataset, MH $p_{V_{in}}(u)$, $q_{S}(u, v)$ ends up with the highest F1 score of 0.823 (SVM). MHs and Role2Vec outperform Node2vec at walk number of 1. As the walk number escalates to 3, the performance gap narrows.
On the ETC dataset, the MH variant $p_{D_{in}}(u)$, $q_{S}(u, v)$ achieves an F1 score of 0.938 (LR) and 0.909 (SVM) with walk number 15, both scores being the highest performed.
In scenario of Bitcoin Elliptic dataset with a walk number of 10, MH $p_{D_{in}}(u)$, $q_{E}(u, v)$ reaches an F1 score of 0.780 (SVM) surpassing other algorithms.
Since the bitcoin elliptic dataset lacks the amount of transfers for each edge due to intellectual property issues (no exact description for features provided), $p_{V_{in}}(u)$ is missing, and $p_{V/D_{in}}(u)$, $q_{E}(u, v)$ here mean $p_{D_{in}}(u)$ for the bitcoin elliptic dataset, while for other datasets, they mean $p_{V_{in}}(u)$.
For the Phishing Node dataset, with a walk number of 8, MH $p_{V_{in}}(u)$, $q_{S}(u, v)$ obtains the highest F1 score of 0.802 (SVM).
Taking a comprehensive view of results, MH algorithms demonstrate improved performance with the same constrained walk lengths compared to other algorithms, though Role2Vec performs consistently competitively compared to MHs.
As Figrue \ref{fig:mean_walk_length} presents, MH's de facto mean walk lengths exceed other algorithms in Ethereum and Bitcoin Elliptic dataset. 
The random walk process may reach a node with no neighbors (a sink node), resulting in a walk length shorter than the preset walk length.
The mean walk lengths demonstrate MH's ability to leap-walk to nodes less likely to be sink nodes, thereby avoiding sparsely connected regions.
MH's efficiency is demonstrated by the ability to achieve comparable performance with fewer walks and with an increased walk length (within a fixed preset limit) compared to the other methods in the experiment.

\subsection{Discussion}
Incorporating the Unbiased Update with the Metropolis-Hastings approaches, presents our proposed strategy for general node representation learning on blockchain transaction networks towards addressing temporality and scalability. 
%
%
The fusion will be studied in the future work.
The Unbiased Update allows efficient updates to vertex representations following changes in the graph, preventing from re-training the whole graph from scratch when the transaction graph keeps growing. 
On the other hand, MH emphasizes performing leap-walk to the significant nodes for a node sampling process, thus reducing the number of walks required for achieving the expected performance in the downstream tasks. 

\subsubsection{Potential Applications}

\begin{description}[leftmargin=0pt, itemsep=0.5em, font=\normalfont\textbf]

\item[Transaction Networks Monitor]
Blockchain transactions are continuously added to the newest block, forming a dynamic graph of interactions.
After a specific interval, a snapshot of the graph back to certain date up to the newest transaction would be formed. 
The Unbiased Update can swiftly adjust vertex representations in response to these additions, keeping the model current with the newest addresses (nodes) to ensure they have embeddings.
Simultaneously, the MH approach can be employed to efficiently sample significant or suspicious nodes, aiding in real-time fraud, abnormal node detection. 
This fusion is valuable for blockchain analysts to build a pipeline that focuses on an incremental node representation learning (via Unbiased Update) and relevant importance (via MH leap-walk) to efficiently find insights or detect irregular nodes.

\item[Smart Contract Mapping]
Concentrating on the nodes representing smart contracts, a specialized node embedding space for smart contracts can be obtained. 
In this space, the contracts with analogous functionalities or frequent interactions are spatially proximate. 
Leveraging proper stationary distributions of the proposed fused approach will help in efficient identification of specific categories of smart contracts. 
The detection of smart contracts for the shared functionalities or common interaction patterns can effectively facilitate a categorization of smart contracts.

\end{description}

\subsubsection{Limitation}

\

\noindent While Metropolis-Hastings with Unbiased Update offers benefits in requiring fewer walks to achieve specific node embedding performance measures and supports incremental learning, if the proposal distribution $q$ considers $h$-hop, it introduces the extra complexity compared to the methods such as Node2Vec.
For Node2Vec, the complexity of generating walks is $O(n \times l \times d)$. In contrast, the MH approaches have an overall complexity of $O(n \times l \times d')$, where $d'$ denotes the average nodes within $h$-hops (corresponds to $N_h(curr)$ in Algorithm. \ref{algo:MH_sampling}). 
Despite MH's potential in minimizing walk counts, it's essential to restrict $d'$ to ensure that the complexity introduced during the walk generation doesn't negate the efficiency gains from fewer walks in the subsequent skip-gram training.
$h$ should be limited according to the graph. 
Furthermore, when the number of $h$-hop nodes is substantial, random sampling can be employed to analyze a subset (such as limit $d'$ to be at most $h*d$), and or cache the $h$-hop neighbors for each node after it has been computed once.

\section{Related work}
Academic studies related to our work mostly are pertaining to node representation learning for blockchain transaction networks.
Martin et al. provided an overview of fundamental machine learning techniques for deriving node embeddings aimed at anomaly detection \cite{martin2022anomaly}, a prevalent objective in this area of research.
Other objectives include, but are not limited to, price prediction \cite{liu2019hyperbolic}, entity identification \cite{wang2022heterogeneous}, and transaction forecasting \cite{lin2020modeling}, etc.
Blockchain transactions can be conceptualized either as a static network (graph) or as evolving dynamic networks.
Considering a static network, a wide range of studies focused on the features that enhance the learned node representation. 
Wu et al. introduced a biased random walk method for Ethereum, with a bias in transaction value and timestamp, prioritizing higher values or recent dates \cite{wu2020phishers}.
B{\'e}res et al. applied various random walk-based representation learning algorithms to Ethereum transaction networks to discover multiple addresses that belong to the same user \cite{beres2021blockchain}, finding that Role2Vec \cite{ahmed2018learning} as one of the best performing methods.
Liu et al. learned an embedding space of blockchain addresses in a hyperbolic space with higher-hierarchical nodes being more discernible in the space for predicting Ethereum price fluctuations \cite{liu2019hyperbolic}.
In dynamic network research, the focus has been on integrating multiple graph snapshots of a time line as the input for node or edge forecasting.
Cai et al. introduced a temporal graph neural network model to identify anomalous edges by analyzing atypical and time-based sub-graphs \cite{cai2021structural}.
Pareja et al. presented an evolving GCN model which utilizes RNN principles to capture temporal patterns and evaluated it on the Bitcoin OTC dataset\footnote{https://snap.stanford.edu/data/soc-sign-bitcoin-otc.html} \cite{pareja2020evolvegcn}.
Khan et al. discussed and emphasized future research directions of graph-based mining of blockchain transactions,  stated that given the dynamic nature of accounts and transactions, the utilized ML models should address data drifts. Techniques like incremental learning, machine unlearning, and continuous learning are beneficial \cite{khan2022graph}.
A recent review by Zhou et al. \cite{zhou2023cryptocurrency} of graph representation learning on blockchain transaction echoed a similar perspective.

\section{Conclusion}
We present a graph representation learning approach for obtaining node embeddings in blockchain transaction networks, which combines Metropolis-Hastings sampling with an Unbiased Update towards addressing evolving network dynamics and scalability challenges.
The approach incorporates incremental learning to mitigate re-training on the whole transaction graph.
The proposed Metropolis-Hastings random walk-based approach performs leap-walk towards significant nodes, reducing occurrences of walk to sparsely connected regions. 
The method seeks to generate fewer yet more significant walks, achieving comparable performance. By reducing the number of walks, it lowers the computational overhead of skip-gram training, enhancing the efficiency of node representation learning.
The experiments showed the improved node classification results across various blockchain transaction datasets using fewer walks compared to the other methods.

\bibliographystyle{ACM-Reference-Format}
\bibliography{sample-base}

\end{document}